\documentclass[aps,prd,12pt,showpacs,notitlepage,nofootinbib,tightenlines]{revtex4-1}
\usepackage{amsmath}
\usepackage{bm}
\usepackage{times}
\usepackage{braket}
\usepackage{color}
\usepackage{epsfig}
\usepackage{slashed}
\usepackage{hyperref}
\usepackage{multirow}
\newcommand{\beq}{\begin{eqnarray}}
\newcommand{\eeq}{\end{eqnarray}}

\newcommand{\ord}{{\cal O}}
\newcommand{\cala}{{\cal A}}
\newcommand{\calb}{{\cal B}}

\def \cpc{ {\bf Chin. Phys. C} }

\def \ijmpa{ {\bf Int. J. Mod. Phys. A}  }
\def \copc{ {\bf Comput. Phys. Commum. } }
\def \epjc{{\bf Eur. Phys. J. C} }

\def \npb{ {\bf Nucl. Phys. B} }
\def \plb{ {\bf Phys. Lett. B} }

\def \prt{  {\bf Phys. Rept.} }

\def \prd{ {\bf Phys. Rev. D} }
\def \prl{ {\bf Phys. Rev. Lett.}  }

\def \zpc{ {\bf Z. Phys. C}  }
\def \jhep{ {\bf J. High Energy Phys.}  }

\def \cpl{ {\bf Chin. Phys. Lett.}  }

\definecolor{Red}{rgb}{1.,0.,0.}

\definecolor{Blue}{rgb}{0.,0.,1.}

\definecolor{nicered}{rgb}{0.7,0.1,0.1}
\definecolor{nicegreen}{rgb}{0.1,0.5,0.1}
\def\lsim{ {\ \lower-1.2pt\vbox{\hbox{\rlap{$<$}\lower6pt\vbox{\hbox{$\sim$}}}}\ } }
\def\gsim{ {\ \lower-1.2pt\vbox{\hbox{\rlap{$>$}\lower6pt\vbox{\hbox{$\sim$}}}}\ } }

\hypersetup{colorlinks,citecolor=nicegreen,linkcolor=nicered}
\begin{document}
\title{Search for single production of the heavy vectorlike $T$ quark with $T\to th$ and $h\to \gamma\gamma$ at the high-luminosity LHC}
\author{Yao-Bei Liu\footnote{E-mail: liuyaobei@sina.com}}
\affiliation{Henan Institute of Science and Technology, Xinxiang 453003, People's Republic of China}

\begin{abstract}
The vectorlike top partners $T$ are predicted in many extensions of the Standard Model (SM). In a simplified model including a single vectorlike $T$ quark with charge $2/3$, we investigate the process $pp \to Tj$ induced by the couplings between the top partner with the first and the third generation quarks at the LHC. We find that the mixing with the first generation can enhance the production cross section. We further study the observability of the single heavy top partner through the process $pp \to T(\to th)j\to t(\to b \ell \nu_{\ell})h( \to \gamma\gamma) j$ at the high-luminosity~(HL)-LHC~(a 14 TeV $pp$ collider with an integrated luminosity of 3 ab$^{-1}$).
  For three typical heavy $T$ quark masses $m_T=600,800$ and 1000 GeV, the $3\sigma$ exclusion limits, as well as the $5\sigma$ discovery reach in the parameter plane of the two variables  $g^{\ast}-R_L$, are respectively obtained at the HL-LHC.
\end{abstract}


\maketitle

\section{Introduction}
The discovery of a 125 GeV Higgs boson at the CERN LHC \cite{atlas,cms} has heralded the
beginning of a new era of Higgs physics. However, it is theoretically difficult to understand why the Higgs boson has a mass of $\sim$100 GeV. In order to offer a potential solution to the hierarchy problem, many new physics theories beyond the Standard Model (SM) predict the existence of new heavy fermions, which can stabilize the Higgs boson mass and protect it from dangerous quadratic divergences~\cite{h1,h2}. In many cases, such as little Higgs models~\cite{littlehiggs}, extra dimensions~(ED) \cite{ED}, composite Higgs models~\cite{Agashe:2004rs} and twin Higgs models \cite{twinhiggs} etc., these new heavy fermions are heavy vectorlike top partners, which have
the same spin and only differ in the embedding into representations of the weak isospin, $SU(2)_{L}$.
The phenomenology of new heavy quarks has been widely
studied in literatures; see for
example~\cite{p1,p2,p-lh1,p-lh2,p-cp1,p-cp2,p-twin,p5,p6,p7,p8}.

Very recently, the ATLAS and CMS Collaborations have performed the searches for $T$ quarks and have placed limits on its mass
ranging from $720$ to $950$ GeV for different $T$ quark branching fractions ($bW$, $tZ$ and $th$)~\cite{atlas-1,cms-1}. Most of these experimental searches assume
the $T$ quarks to be pair produced via the strong interaction, and these bounds strongly depend on the assumptions on the decay
branching ratios and the properties of the top partner. Apart from direct searches, the indirect searches for the top partners through their contributions to the electroweak precision observables~\cite{ew1,ew2}, $Z$-pole observables~\cite{rb,prd88-094010} and the Higgs decay channels in the various final states \cite{higgs-decay1,higgs-decay2,higgs-decay3} have been extensively investigated. As pointed out in Refs.~\cite{s1,s2,s3,s4,s5}, single production of top partners starts to dominate over pair production for high $M_T$ (about $M_T \gsim 1$ TeV) due to larger phase space. Very recently, the analyses of the singly produced top partners that decay to $bW$ and $tZ$ have been performed in Refs.~\cite{tp-single-1,jhep2015-01-088}. Furthermore, the authors in Ref.~\cite{jhep1604-014} studied the search strategies of the single top partner production with all the possible decay modes~($tZ$, $th$ and $Wb$) by using the boosted object tagging methods. As a result, several ATLAS and CMS
vectorlike quark searches also target their single production
mode~\cite{atlas-single,cms-single}. However, most of the studies are based on the assumption that the $T$ quarks only couple to the first generation of quarks or the third generation $b$ quark.

Considering the constraints from flavor physics~\cite{bound1,bound2,bound3,jhep2014-09-130,Buchkremer:2013bha,jhep2015-09-012},
the top partners can mix in a sizable way with
lighter quarks, which could have a
severe impact on electroweak vectorlike quark processes
at the LHC~\cite{jhep1108-080}. Using the cut-and-count analysis, the authors in Ref.~\cite{jhep2015-02-032} studied the LHC discovery potential of the $T\to tZ$ channel in the trilepton decay mode in single production, for a singlet $T$
quark mixing with the first generation. Such mixing can enhance the single production due
to the presence of valence quarks in the initial state. In this work, we study the single vectorlike $T$ quark production through the process $pp \to T(\to th)j\to t(\to b \ell \nu_{\ell})h( \to \gamma\gamma) j$ in a simplified  model, as
presented in Ref.~\cite{Buchkremer:2013bha}. The benefit of using the simplified effective  theory  is that the results of the studies could be used to make predictions for more complex models including various types of top partners~(see, e.g.,~Refs.~\cite{jhep2015-02-032,jhep2013-10-160,jhep1407-142,jhep1412-080,jhep1412-097,jhep1409-060}).
In comparison with the existing searches for other decay modes, the $h \to \gamma\gamma$ channel has a small cross section but has the great
advantage that most QCD backgrounds are gone, such as done in Refs.~\cite{wulei,liu}.  This has recently been done by the CMS Collaboration in
events with at least one top partner undergoing the $T\to th(h\to \gamma\gamma)$ decay chain from the $T\bar{T}$ production~\cite{cms-haa}, which has excluded the existence of
top quark partners with mass up to 540 GeV using 19.7 fb$^{-1}$ of integrated luminosity under the hypothesis that $Br(T\to th)=100\%$.
The rough estimation in Ref.~\cite{jhep1604-014} has shown that it is challenging to observe the $T$ quark with an $h_{\gamma\gamma}$ signal at the 14 TeV with 100 fb$^{-1}$ due to the small production cross section.
Thus, here we consider the potential for the HL-LHC with an integrated luminosity of 3 ab$^{-1}$. We expect that such a channel may become a complementary to other channels in the searches for the heavy vectorlike top partner.

This paper is arranged as follows. In Sec. II, we briefly describe the simplified model and calculate the single top partner production cross section involving the mixing with both the first and third generate quarks. In Sec.
III, we discuss the observability of the top partner through the process $pp \to T(\to th)j\to t(\to b \ell \nu_{\ell})h( \to \gamma\gamma) j$
at the HL-LHC. Finally, a short summary is given in Sec. IV.

\section{Top partner in the simplified model}
\subsection{Brief review of the simplified model}
To capture all the essential features of the new heavy top partners while remaining as model
independent as possible, the authors of Ref.~\cite{Buchkremer:2013bha} have proposed a generic parametrization of an effective Lagrangian for top
partners with different electromagnetic charge, where they considered vectorlike quarks embedded in general
representations of the weak $SU(2)$ group. Here we mainly consider the case in which the top partner is an $SU(2)$ singlet and can mix and decay directly into the first and third generation quarks. The interactions of the vectorlike $T$ quark with gluons and photons are governed by the gauge
symmetries.

The simple
Lagrangian that parametrizes the $T$ couplings to quarks and electroweak bosons is
\cite{Buchkremer:2013bha}
\begin{eqnarray}
{\cal L}_{T} =&& \frac{gg^{\ast}}{2}\left\{ \frac{S_R}{\sqrt{2}} \right.[\bar{T}_{L}W_{\mu}^{+}
    \gamma^{\mu} d_{L}]  + \frac{C_{R}}{\sqrt{2}} [\bar{T}_{L} W_{\mu}^{+} \gamma^{\mu} b_{L}]\nonumber \\
  && + \frac{S_{R}}{2\cos \theta_W} [\bar{T}_{L} Z_{\mu}^{+} \gamma^{\mu} u_{L}] +\frac{C_{R}}{2\cos \theta_W} [\bar{T}_{L} Z_{\mu}^{+} \gamma^{\mu} t_{L}] \nonumber \\
  &&  \left. - S_R\frac{M_{T}}{v}[\bar{T}_{R}hu_{L}] - C_R\frac{M_{T}}{v}[\bar{T}_{R}ht_{L}]- C_R\frac{m_{t}}{v} [\bar{T}_{L}ht_{R}] \right\}+ h.c. ,
  \label{TsingletVL}
\end{eqnarray}
where
\beq
S_R=\sqrt{\frac{R_L}{1+R_L}},\quad  C_R=\sqrt{\frac{1}{1+R_L}}.
\eeq
In Eq.~\eqref{TsingletVL}, $g$ is the $SU(2)_L$ gauge coupling constant, $\theta_W$ is the Weinberg angle, $v\simeq 246$ GeV, and  the subscripts $L$ and $R$ label the chiralities of the fermions. Besides the top partner mass $m_T$, there are two free parameters:
\begin{itemize}
\item
 $g^{\ast}$, the coupling
strength to SM quarks in units of standard couplings, which is only
relevant in single production. In general, the value of $g^{\ast}$ can be taken in the range $0.1- 0.5$~\cite{jhep2015-01-088}. A more complete description of the limits on the mixing with the vectorlike quarks has been studied in Ref.~\cite{prd88-094010}.
\item
$R_L$, the generation mixing coupling, which controls the
share of the $T$ coupling between first and third generation quarks. In the extreme case, $R_L = 0$ and $R_L = \infty$, respectively, correspond to
coupling to third generation quarks and first generation of quarks only.
\end{itemize}

For $R_L=0$, the branching fractions of $T$ into $th$, $tZ$ and $bW$ reach a good
approximation\footnote{Here we consider the case in which $m_T\gg m_t$, as shown in Ref.~\cite{jhep2015-01-088}.}, given by the ratios $1 : 1 : 2$ as expected by the Goldstone boson equivalence theorem for a heavy singlet
top partner~\cite{s1}. It should be mentioned that not only can the nonvanishing $R_L$ alter the branching ratios, but it can also alter the case when $g^{\ast}$ gets close to 1~(see, e.g., \cite{prd88-094010}).  A full study of the precision bounds
of this particular model is beyond the scope of this paper, as we only use this model as illustration for top partner search strategies. As shown in Refs~\cite{bound1,bound2,bound3,jhep2014-09-130,Buchkremer:2013bha,jhep2015-09-012,jhep1108-080}, these parameters can be constrained by the flavor physics and the oblique parameters. Here we take a conservative range for the mixing parameter: $0\leq R_L\leq 2$.
For some details about this effective Lagrangian in terms of complete
models, one can see Refs. \cite{jhep2015-01-088,jhep2014-09-130,Buchkremer:2013bha,jhep2015-09-012}.

\subsection{Single production of top partner $T$}
\begin{figure}[htb]
\begin{center}
\vspace{-1.0cm}
\centerline{\epsfxsize=14cm \epsffile{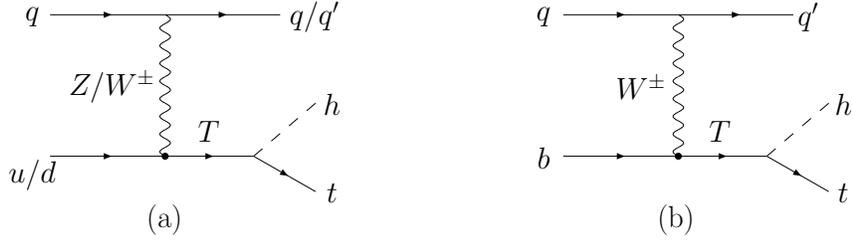}}
\vspace{-14.5cm}
\caption{Feynman diagrams for the process $pp \to Tj\to thj$ at the
LHC via couplings of the $T$ to (a) first generation quarks and (b) third generation quarks.}
\label{fey}
\end{center}
\end{figure}
At the LHC, the top partner can be singly produced in association with a light jet. In this paper we mainly study the LHC discovery potential of the $T$ quark from the decay channel $T\to th$. The Feynman diagrams for the process $pp\to Tj\to thj$ are plotted in Fig.~\ref{fey}, where the top partner is produced due to the interaction with
light quarks or due to the interaction with the $b$ quark. From the Lagrangian in
Eq.~\eqref{TsingletVL}, we know that the production cross sections coming from these two sets of diagrams are very sensitive to the mixing coupling parameter $R_L$.

The model file of the singlet $T$ quark \cite{VLQweb} is implemented via the FeynRules package \cite{feynrules}. The cross sections are calculated at tree level using MadGraph5-aMC$@$NLO \cite{mg5} and checked by CalcHEP~\cite{calchep}. We use CTEQ6L as the
parton distribution function~\cite{cteq} and set the renormalization scale $\mu_R$ and factorization
scale $\mu_F$ to be $\mu_R=\mu_F=(m_T)/2$.  The SM
input parameters are taken as follows \cite{pdg}:
\begin{align}
m_H&=125{\rm ~GeV}, \quad m_t=173.21{\rm ~GeV}, \quad m_W=80.385{\rm ~GeV},\\ \nonumber
\alpha(m_Z)&=1/127.9, \quad \alpha_{s}(m_Z)=0.1185, \quad G_F=1.166370\times 10^{-5}\ {\rm GeV^{-2}}.
\end{align}

\begin{figure}[thb]
\begin{center}
\vspace{-0.5cm}
\centerline{\epsfxsize=10cm \epsffile{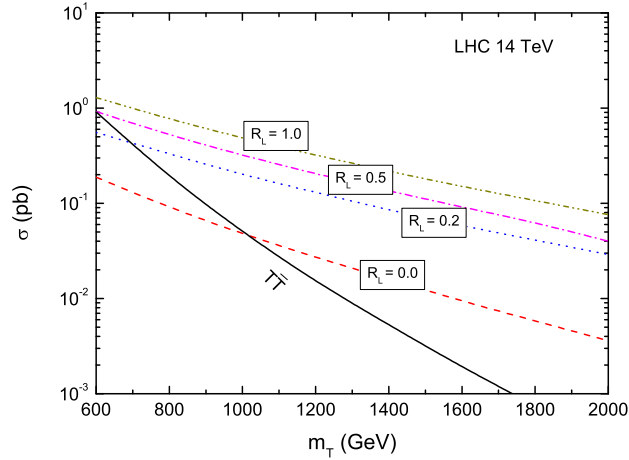}}
\caption{The dependence of the cross sections $\sigma$ for the process $pp\to Tj+\bar{T}j$ on the $T$ quark mass
$m_T$ at the 14 TeV LHC with $g^{\ast}=0.1$ and several values of $R_L$. The solid curves show the $T\bar{T}$ pair production cross section.}
\label{cross1}
\end{center}
\end{figure}

In Fig.~\ref{cross1}, we show the dependence of the cross sections $\sigma$ for the process $pp\to Tj+\bar{T}j$ on the $T$ quark mass
$m_T$ at the 14 TeV LHC for $g^{\ast}=0.1$ and several values of $R_L$. For comparison, we also show the $T\bar{T}$ pair production cross section. One can see that, even without mixing~(in the case of $R_L=0$), the single production of top partners starts to dominate
over pair production for $m_T\gtrsim 1$ TeV. On the other hand, the values of for the single production cross sections are very sensitive to $R_L$. This implies that the mixing with the first generation can enhance the single production, especially due
to the presence of valence quarks in the initial state.

\begin{figure}[thb]
\begin{center}
\vspace{-0.5cm}
\centerline{\epsfxsize=9cm \epsffile{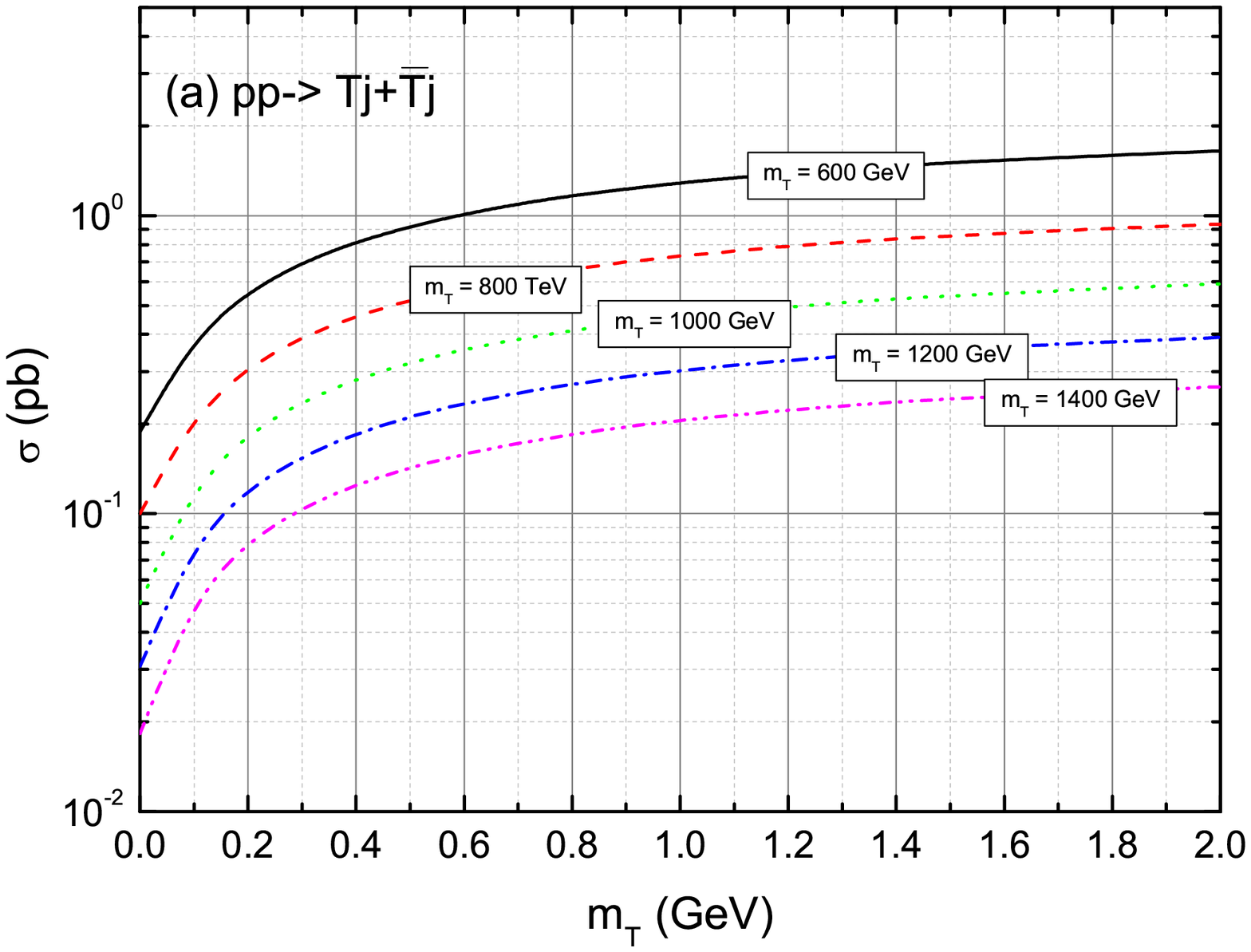}\epsfxsize=9cm \epsffile{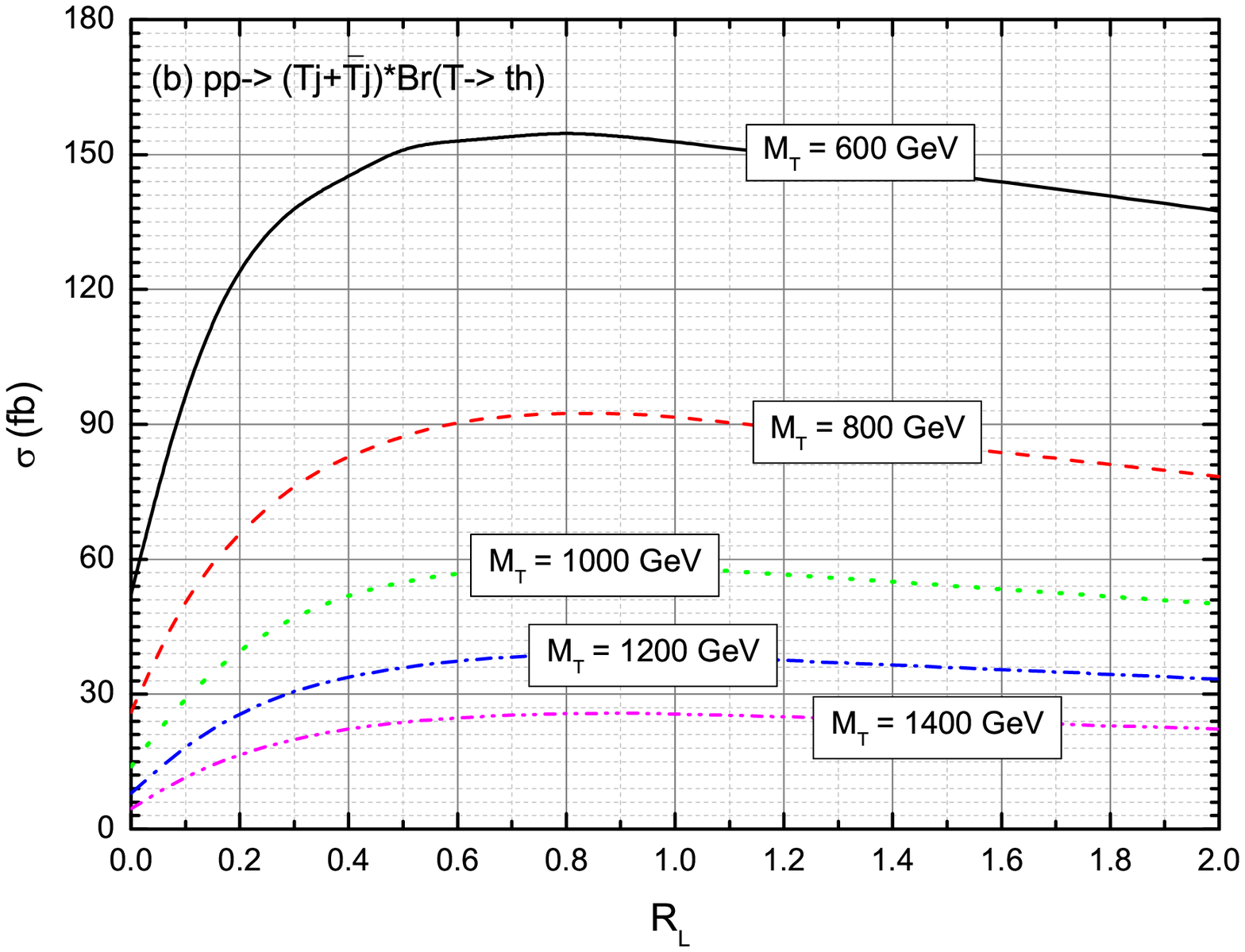}}
\caption{The dependence of the cross sections $\sigma$ on the mixing parameter $R_L$ at 14 TeV LHC for the processes (a)$pp\to (Tj+\bar{T}j)$, and (b) $pp\to (Tj+\bar{T}j)\to (t+\bar{t})hj$. Here we take $g^{\ast}=0.1$ and five typical $T$ quark masses $m_H=600, 800, 1000, 1200$ and 1400 GeV, respectively.  }
\label{cross2}
\end{center}
\end{figure}

In Figs.~3(a) and 3(b), we show the dependence of the cross sections  $\sigma$  on the mixing parameter $R_L$
at the 14 TeV LHC for the processes $pp\to (Tj+\bar{T}j)$ and $pp\to (Tj+\bar{T}j)\to (t+\bar{t})hj$, respectively. We generate  five benchmark points varying the $T$ quark mass in steps of 200 GeV in the range [600;~1400] GeV with $g^{\ast}=0.1$.
One can see that (i) in the range of $R_L<1$, the production cross sections for the processes $pp\to (Tj+\bar{T}j)$ and $pp\to (Tj+\bar{T}j)\to (t+\bar{t})hj$ both increase largely with the increase of $R_L$. (ii) For $R_L>1$, the production cross section for the process $pp\to (Tj+\bar{T}j)$ will become slightly large with the increase of $R_L$, while the production cross section for the process $pp\to (Tj+\bar{T}j)\to (t+\bar{t})hj$ will become small with the increase of $R_L$. This effect is mainly due to the increased admixture of valence quarks in production, mitigated by a reduced $T\to th$ branching ratio with increasing $R_L$. For the process $pp\to (Tj+\bar{T}j)\to (t+\bar{t})hj$,
the
cross section will reach a maximum for $R_L\simeq 1$, which corresponds
to $50\%-50\%$ mixing.

\section{LHC observability of $T(\to th)j\to t(\to b\ell^{+} \nu_{\ell})h(\to \gamma\gamma)j$}
In this section, we perform the
Monte Carlo simulation and explore the sensitivity of single top partner at 14 TeV LHC through the channel,
\begin{equation}\label{signal}
pp \to T(\to th)j\to t(\to b \ell \nu_{\ell})h( \to \gamma\gamma) j.
\end{equation}
The corresponding free parameters are the top partner mass $m_T$, the
coupling $g^{\ast}$ which governs the top partner single production, and
the mixing parameter $R_L$. We take three typical values of the $T$ quark mass: $m_T=600,800,1000$ GeV with $g^{\ast}=0.1$ and $R_L=0.5$. Throughout this analysis, we assume that the presence of a $T$ quark in the $h\to \gamma\gamma$ loop would not change the SM branching ratio value of $0.23\%$. Obviously, the cross sections are proportional to $(g^{\ast})^{2}$. The QCD next-to-leading order~(NLO) prediction for single productions is calculated in Refs.~\cite{jhep2015-01-088,1610.04622}. Following Ref.~\cite{jhep2015-01-088}, here we take the conservative value of the $K$-factor as 1.14 for the signal.

Signal and background events are generated at leading order using
MadGraph5-aMC$@$NLO.  PYTHIA \cite{pythia} and Delphes \cite{delphes} are used to perform the parton
shower and the fast detector
simulations, respectively. When generating the parton-level events,
we assume $\mu_R=\mu_F$ to be the default event-by-event value. The anti-$k_{t}$ algorithm
\cite{antikt} with parameter $\Delta R=0.4$ is used to reconstruct jets. Lastly, the program of MadAnalysis5 \cite{ma5} is used for analysis, where the (mis)tagging efficiencies
and fake rates are assumed to be their default values.

The main SM backgrounds which yield final states identical to the signal include two parts: the
resonant and the nonresonant backgrounds. For the former, they mainly come from the processes that
have a Higgs boson decaying to diphoton in the final states, such as $thj$ and $t\bar{t}h$ productions. For the latter, the main background processes contain the diphoton events
 produced in association with the top quarks, such as $tj\gamma\gamma$ and $t\bar{t}\gamma\gamma$ production. Besides, with fake photons due to misidentified jets or electrons, the reducible backgrounds such as $tjj\gamma$, $t\bar{t}\gamma$ and $t\bar{t}\gamma j$ can also be the sources of backgrounds for our signal. However, we have not included all these potentially dangerous contributions in the analysis due to very low keeping efficiency (at the order of $10^{-7}$) after applying the suitable cuts.
The MLM matching scheme is used, allowing up to
four additional partons in the matrix element~\cite{MLM}. The cross sections of $thj$, $t\bar{t}h$ and $t\bar{t}\gamma\gamma$ production are normalized to their NLO values~\cite{nlo-thj,nlo-tth,14050301}.

In our simulation, we generate $10^{6}$ events for the signals and the backgrounds, respectively. All events are first subject to basic cuts:
\begin{itemize}
\item
The isolated lepton with transverse momentum $p_{T}^{\ell} > 20 \rm ~GeV$ and $|\eta_{\ell}|<2.5$.
\item
The $b$-tagged jet with transverse momentum $p_{T}^{b} > 25 \rm ~GeV$ and $|\eta_{b}|<2.5$.
\item
The light jets with $p_{T}^{j} > 25 \rm ~GeV$ and $|\eta_{j}|<5$.
\item
The photons with $p_{T}^{\gamma} > 20 \rm ~GeV$ and $|\eta_{\gamma}|<2.5$.
\end{itemize}
For the signal, we require exactly one charged lepton, one light jet, one $b$-jet and two photons in the final state. To trigger the signal events, $N(\ell)=1$, $N(b)=1$ and $N(j) \leq 2$ are applied, which can help to suppress the background events effectively, especially to the events with fake particles.

\begin{figure}[htb]
\begin{center}
\centerline{\epsfxsize=8cm\epsffile{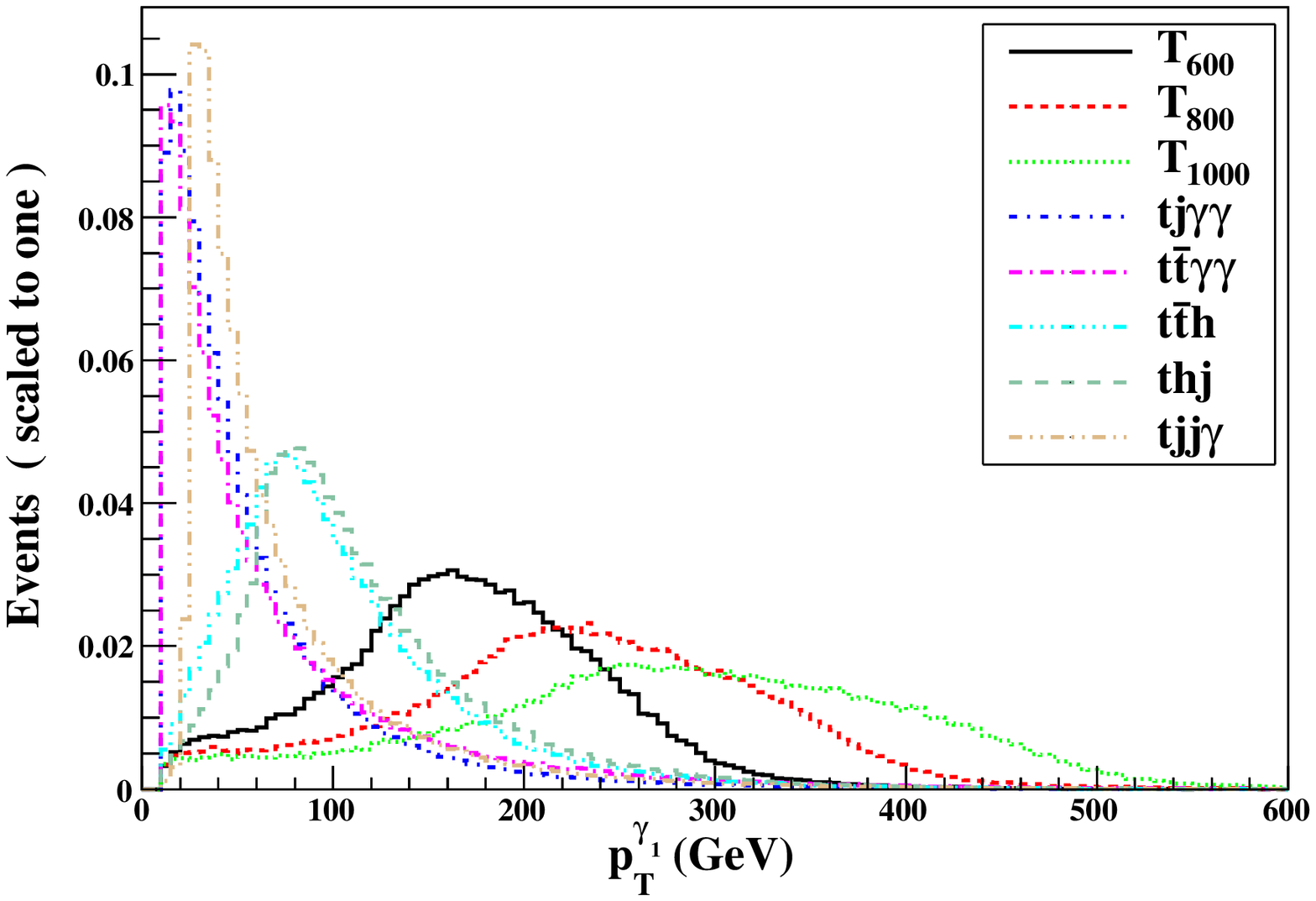}
\hspace{-0.5cm}\epsfxsize=8cm\epsffile{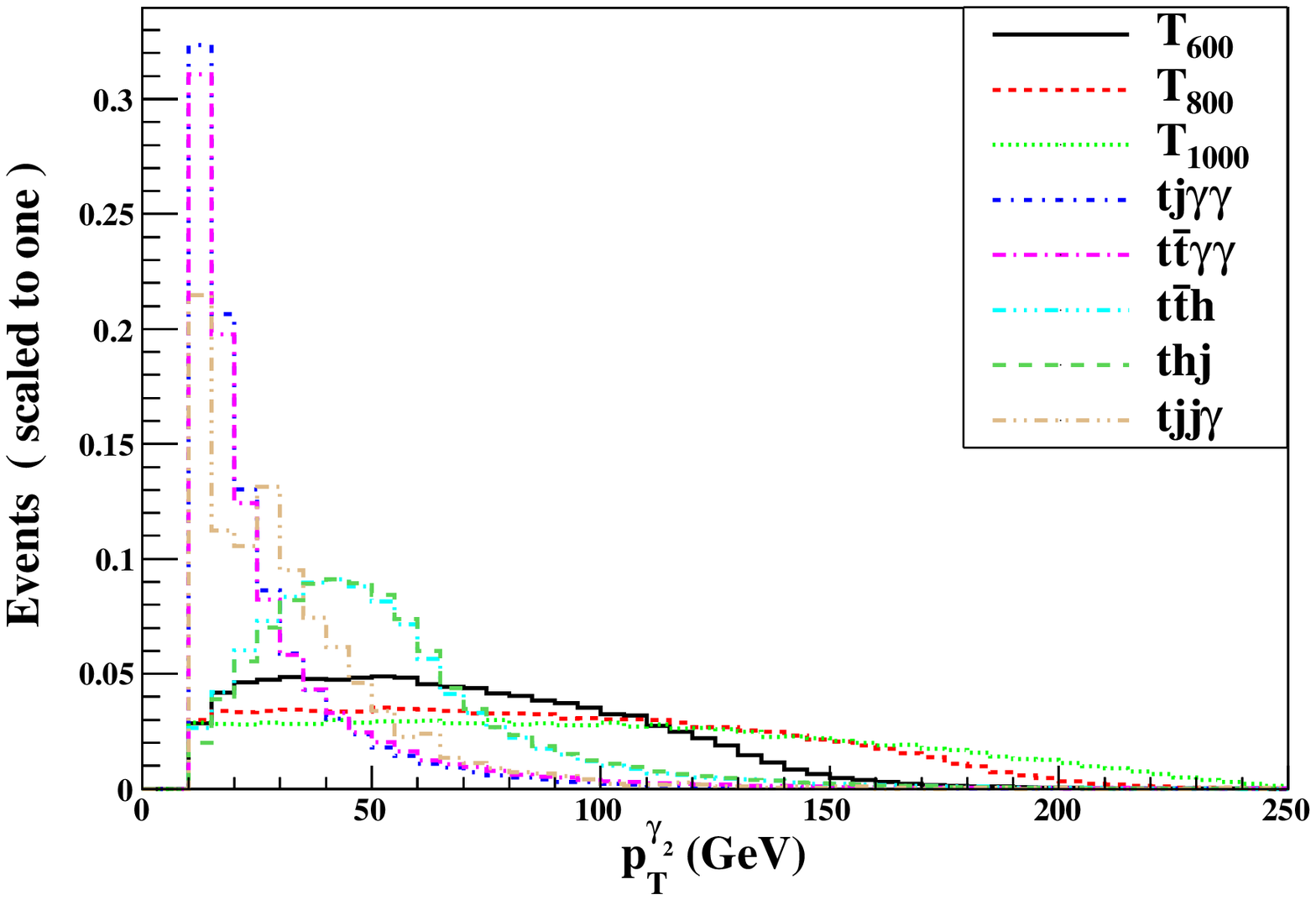}}
\caption{Normalized distributions of transverse momenta $p_{T}^{\gamma_{1}}$ and  $p_{T}^{\gamma_{2}}$
in the signals and backgrounds at 14 TeV LHC.}
\label{photon}
\end{center}
\end{figure}

\begin{figure}[htb]
\begin{center}
\vspace{-0.5cm}
\centerline{\epsfxsize=10cm \epsffile{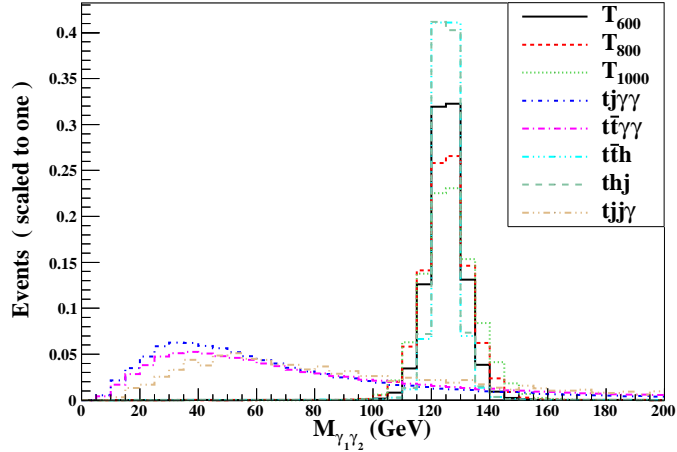}}
\caption{Normalized invariant mass distribution of two photons at 14 TeV LHC.}
\label{higgs}
\end{center}
\end{figure}

In Figs~\ref{photon} and \ref{higgs}, we show the transverse momentum distributions of two photons ($p_{T}^{\gamma_{1}}, p_{T}^{\gamma_{2}}$) and the invariant mass distribution $M_{\gamma\gamma}$ in the signal and backgrounds at 14 TeV LHC. Since the two photons in the signal and the resonant backgrounds come from the Higgs boson, they have the harder $p_T$ spectrum than those in the nonresonant backgrounds.
Thus, we can apply the following cuts to suppress the nonresonant backgrounds:
\beq
p_{T}^{\gamma_1}>120 \rm ~GeV, \quad p^{\gamma_2}_{T}>60 \rm ~GeV.
\eeq
From Fig.~\ref{higgs}, we can see that the signals and the resonant backgrounds, including the Higgs boson, have peaks around 125 GeV. Thus, we can further reduce the nonresonant backgrounds by the following cut:
\beq
120 \rm~GeV <M_{\gamma_{1}\gamma_{2}}<130 \rm ~GeV.
\eeq

\begin{figure}[htb]
\begin{center}
\vspace{-0.5cm}
\centerline{\epsfxsize=10cm \epsffile{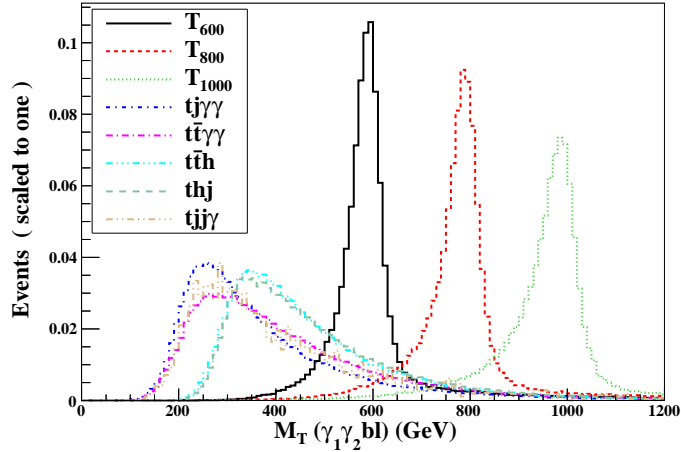}}
\caption{Normalized transverse mass distribution for the $\gamma_{1}\gamma_{2}b \ell \slashed E_T$ system at 14 TeV LHC.}
\label{mt}
\end{center}
\end{figure}

As defined in MadAnalysis5~\cite{epjc-74-3103}, the $T$ quark transverse cluster mass can be reconstructed as
\beq
M_T^{2}\equiv(\sqrt{(\sum_{i=\ell,b,\gamma_{1},\gamma_{2}} p_i^{2}+|\sum_{i=\ell,b,\gamma_{1},\gamma_{2}}\vec{p}_{T,i}|^{2}}+|\slashed p_T| )^{2}-|\sum_{i=\ell,b,\gamma_{1},\gamma_{2}}\vec{p}_{T,i}+\vec{\slashed p}_T|^{2},
\eeq
where $\vec{p}_{T,\ell}$,  $\vec{p}_{T,b}$ and $\vec{p}_{T,\gamma}$ are the transverse momenta of the charged leptons, $b$ quarks and photons, respectively, and $\slashed p_T$ is the missing transverse momentum determined by the negative sum of visible momenta in the transverse direction.
In Fig.~\ref{mt}, we show the transverse mass distribution for the $\gamma_{1}\gamma_{2}b \ell \slashed E_T$  system.
From this figure, we can see that the transverse mass distribution $M_T$ has an endpoint around the mass of top partner. Therefore, we can choose $M_T >500$~GeV to reduce the backgrounds.

For a short summary, we require the events to satisfy the following criteria:
\begin{itemize}
\item[(1)] Basic cut: $p_T^{j,b}>25$ GeV, $p_T^{\ell,\gamma}>20$ GeV, $|\eta_{b,\ell,\gamma}|<2.5$,     and  $|\eta_{j}|<5$.

\item[(2)] Cut 1: the basic cuts plus $N(\ell)=1$, $N(b)=1$ and $N(j) \leq 2$;

\item[(3)] Cut 2: cut 1 plus exactly two photons [$N(\gamma)=2$] with $p_T^{\gamma_{1}}>120 \rm ~GeV$, $p_T^{\gamma_{2}}>60 \rm ~GeV$.

\item[(4)] Cut 3: cut 2 plus the invariant mass of the diphoton pair to be in the range $m_h\pm5$ GeV.

\item[(5)] Cut 4: cut 3 plus $M_T > 500$ GeV.
\end{itemize}

\begin{table}[htb]
\begin{center}
\caption{The cut flow of the cross sections (in $10^{-3}$ fb) for the signal and backgrounds ( $tj\gamma\gamma$, $t\bar{t}\gamma\gamma$, $t\bar{t}h$, $thj$ and $tjj\gamma$ )
at the 14 TeV LHC. Here we take the parameters $g^{\ast}=0.1$ and $R_L=0.5$. \label{cutflow}}
\vspace{0.2cm}
\begin{tabular}{|c|c|c|c|c|c|c|c|c|c|}
\hline
\multirow{2}{*}{Cuts}& \multicolumn{3}{c|}{signal for $Tj$}&\multirow{2}{*}{$t\bar{t}h$} &\multirow{2}{*}{$thj$} &\multirow{2}{*}{$tj\gamma\gamma$} &\multirow{2}{*}{$t\bar{t}\gamma\gamma$} &\multirow{2}{*}{$tjj\gamma$} \\ \cline{2-4}
 &{600~GeV}&{800~GeV}&{1000~GeV} &&&&& \\  \cline{1-9}
\hline
Basic cuts& 25.9&15.4&9.57&37.8&4.2&396&184&$21.5\times 10^{3}$ \\ \hline
Cut 1&6.84&3.92&2.24&0.54&0.08&236&4.6&$2.3\times 10^{3}$\\ \hline
Cut 2& 3.15&2.38&1.47&0.067&0.075&4.13&0.15&0.63\\ \hline
Cut 3& 1.73&1.07&0.59&0.038&0.042&0.074&0.003&0.008\\ \hline
Cut 4& 1.69&1.07&0.59&0.015&0.012&0.01&0.001&0.002 \\ \hline
\end{tabular} \end{center}\end{table}

The cross sections of the signal and backgrounds after imposing
the cuts are summarized in Table~\ref{cutflow}.
From Table~\ref{cutflow},
one can see that the jet multiplicity selection $N(j)\leq 2$ (i.e., cut 1) can efficiently suppress the backgrounds involving $t\bar{t}$. By the requirement of exactly two high $p_T$ photons
(i.e., cut-2), all the backgrounds are greatly removed since the photons in the signal are from the boosted Higgs boson in the heavy $T$ quark decay. Obviously, the nonresonant backgrounds are efficiently reduced by
$\ord(10^{-2})$ due to the Higgs mass cut (i.e., cut-3). Thus,
all the backgrounds are suppressed very efficiently after imposing all the selections.

To estimate the observability quantitatively, we adopt the significance measurement~\cite{ss}
\beq
SS=\sqrt{2L\left [ (S+B)\ln\left(1+\frac{S}{B}\right )-S\right ]},
\eeq
where $S$ and $B$ are the signal and background cross sections and $L$ is the integrated luminosity.
Here we define the discovery significance as $SS=5$ and exclusion limits as $SS=3$.
From Table~\ref{cutflow}, we see that the cross sections for the signal are only at the level of $10^{-3}$ fb and thus we take a high integrated luminosity of 3 ab$^{-1}$ at the 14 TeV LHC. The cross section for an arbitrary value of $R_L$ is calculated following the method in Ref.~\cite{jhep2015-02-032}, which has been presented in Appendix A.
\begin{figure}[htb]
\begin{center}
\vspace{1.5cm}
\centerline{\epsfxsize=7cm \epsffile{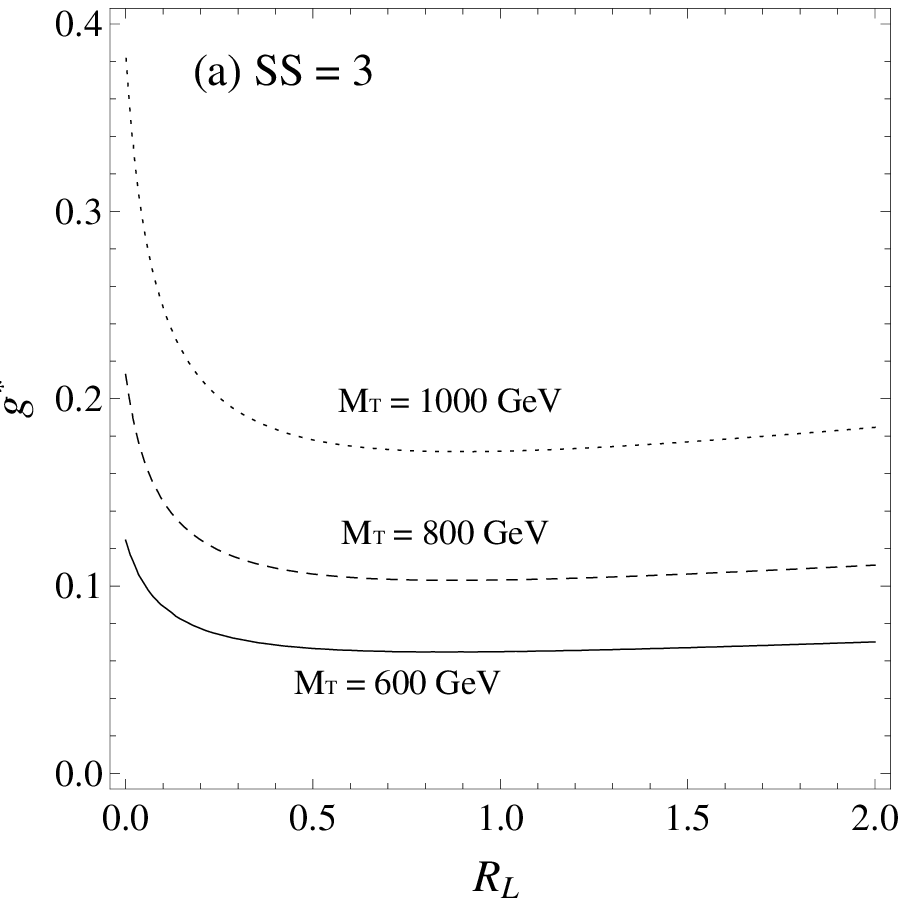}\epsfxsize=7cm \epsffile{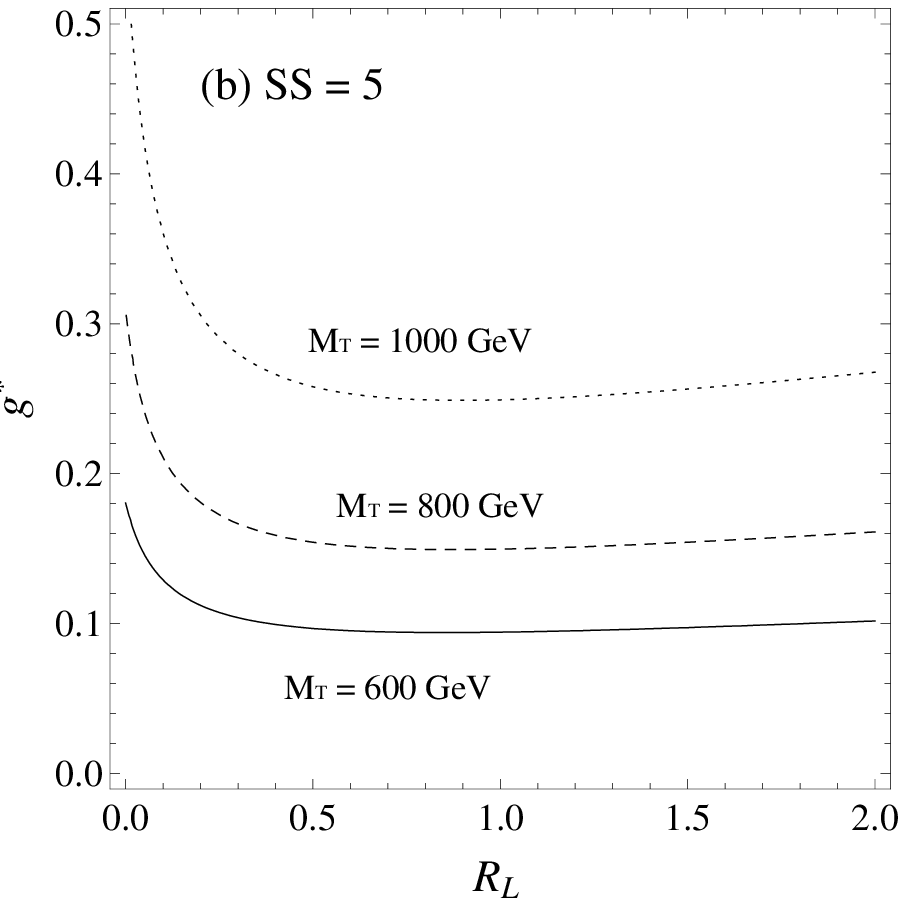}}
\caption{The $3\sigma$ (left) and $5\sigma$ (right) contour plots for the signal in $g^{\ast}-R_L$ at 14 TeV LHC with 3 ab$^{-1}$ of integrated luminosity. }
\label{ss}
\end{center}
\end{figure}

In Fig.~\ref{ss}, we plot the excluded $3\sigma$ and $5\sigma$ discovery reaches in the plane of $g^{\ast}-R_L$ for three fixed typical $T$ masses at 14 TeV LHC with 3 $ab^{-1}$ of integrated luminosity. From Fig.~\ref{ss}, one can see that the $5\sigma$ level discovery sensitivity of $g^{\ast}$ is about 0.18~(0.31) for $m_T=600~(800)$ GeV and $R_L=0$, and it changes as $0.1~(0.16)$ for the nonvanishing $R_L$. As mentioned before, the cross section for the final state (and hence the LHC) reaches a maximum for $R_L\simeq 1$ due to the mixing effects. On the other hand, from the $3\sigma$ exclusion limits one can see that the upper limits on the size of $g^{\ast}$ are given as $g^{\ast}\leq 0.12~(0.21)$ for $m_H=600~(800)$ GeV and $R_L=0$ and that they change as $g^{\ast}\leq 0.08~(0.11)$ for the nonvanishing $R_L$.

\section{CONCLUSION}
The new heavy vectorlike top partner of charge 2/3 appears in many new physics models beyond the SM.
In this paper, we exploited a simplified model with only three free parameters: the top partner mass $m_T$, the electroweak coupling constant $g^{\ast}$ and the generation mixing parameter $R_L$. We first calculated the cross section for the single $T$ production with the decay channel $T\to th$. Then, we investigated the observability of the heavy vectorlike top partner $T$ production through the process $pp\to T(\to th)j\to t(\to b\ell\nu_{\ell})h(\to \gamma\gamma)j$ at the HL-LHC.  The $3\sigma$ exclusion limits, as well as the $5\sigma$ discovery reach in the parameter plane of the two variables  $g^{\ast}-R_L$, are obtained for three typical heavy $T$ quark masses $m_T=600,800$ and 1000 GeV, respectively. For $m_T=600~(800)$ GeV, the upper limits on the size of $g^{\ast}$ are given as $g^{\ast}\leq 0.12~(0.21)$ for $R_L=0$ and $g^{\ast}\leq 0.08~(0.11)$ for the nonvanishing $R_L$.

\begin{acknowledgments}
This work is supported by the Joint Funds
of the National Natural Science Foundation of China (U1304112) and the Foundation of He¡¯nan Educational Committee (2015GGJS-059).
\end{acknowledgments}


\begin{appendix}
\section{Cross section for an arbitrary value of $R_L$}\label{sec:data}
Following the method in Ref.~\cite{jhep2015-02-032}, the cross section for an arbitrary value of $R_L$ is given by\footnote{For a more detailed interpretation of this method, see Sec. 2.1 in Ref.~\cite{jhep2015-02-032}. Here we also consider the case in which the $T$ decay products are much lighter than its mass.}
\beq
\sigma(m_T,R_L)=\sigma(m_T,0.5)\times\frac{\sigma_{pp\to Tj}(m_T,R_L)Br_{T\to th}(m_T,R_L)}{\sigma_{pp\to Tj}(m_T,0.5)Br_{T\to th}(m_T,0.5)}
\eeq
where
\beq
\sigma_{pp\to Tj}(m_T,R_L)&=&\frac{R_L}{1+R_L}\cala_{R_L\to \infty}(m_T)+\frac{1}{1+R_L}\cala_{R_L=0}(m_T), \\
Br_{T\to th}(m_T,R_L)&=&\frac{1}{1+R_L}\calb(m_T).
\eeq
Here $\cala_{R_L\to \infty}$ and $\cala_{R_L=0}$, respectively, represent the production cross section for the heavy $T$ quark due to the interaction to partons belonging to the first and third generations. Note that $\calb(m_T)$ is the $T$ quark branching ratio for its decay into a top quark and a SM Higgs boson, which can be calculated automatically by using MadGraph5-aMC$@$NLO.

Obviously, such cross section also depends on the choice of the PDF and the EW couplings, etc. By considering the ratio of the cross sections evaluated at different values of $R_L$, one can factorize the impact of the above choices. Thus, such a ratio can be used to rescale a given cross section, evaluated by the choice of  $R_L$.
\end{appendix}

\end{document}